\documentclass[preprint,showpacs]{revtex4}
\usepackage{graphicx}
\usepackage{bm}
\usepackage{amsmath} 

\newcommand{\third}{\mbox{${\textstyle \frac{1}{3}}$}}          

\newcommand{\rd}{\textrm{d}}
\begin{document}

\title{Remarks on the analysis  of the reaction $e^+e^-\rightarrow\Sigma^0\bar{\Sigma}^0$}
\date{\today}
\author{G\"oran F\"aldt}\email{goran.faldt@physics.uu.se} 
\affiliation{ Department of physics and astronomy, \
Uppsala University,
 Box 516, S-751 20 Uppsala,Sweden }


\begin{abstract}
We investigate roads for evaluating  model-independent cross-section distributions for 
 the sequential hyperon decay 
$\Sigma^0\rightarrow\Lambda \gamma;\Lambda\rightarrow p\pi^- $ and its corresponding antihyperon  decay.
  The hyperons are produced in the reaction
   $e^+e^-\rightarrow J/{\psi}\rightarrow \Sigma^0\bar{\Sigma}^0$.
Cross-section distributions are calculated using the folding technique. 
\end{abstract}


\maketitle
%

%
%
 \date{\today} 
%

%
%
%
\section{Introduction}\label{ett}

The BESIII experiment \cite{Ablikim17a} is exploring new venues into hyperon physics, based on $e^+e^-$ annihilation into hyperon-antihyperon pairs. In a recent paper \cite{GKa}, 
we investigated in some detail the reaction 
$e^+e^-\rightarrow J/{\psi}\rightarrow \Sigma^0\bar{\Sigma}^0$ 
and its associated  decay chains 
$\Sigma^0\rightarrow\Lambda \gamma;\Lambda\rightarrow p\pi^- $ and 
$\bar{\Sigma}^0\rightarrow\bar{\Lambda} \gamma;\bar{\Lambda}\rightarrow \bar{p}\pi^+ $.
 By measuring this process in the vicinity of the $J/\psi$-vector-charmonium state, one gains information on the strong baryon-antibaryon-decay process 
of the $J/\psi$-vector-charmonium state and also, it offers a model-independent way of measuring weak-decay-asymmetry parameters, that 
in turn can probe {\slshape CP} symmetry \cite{Nature}.

The diagram for the basic reaction $e^+e^-\rightarrow J/\psi \rightarrow \Sigma^0\bar{\Sigma}^0$ is graphed in Fig.1.  Its structure is governed by  two vertices.  
The strength of the lepton-vertex function 
is determined by a single parameter, the electromagnetic-fine-structure 
constant  $\alpha_e$, but 
two complex form factors  $G^\psi_M(s)$ and $G^\psi_E(s)$ are needed for 
the baryonic-vertex function.
 However, we shall not work with the form factors themselves but with certain
 combinations thereof: the strength of form factors $D_\psi(s)$; 
 the ratio of form-factor magnitudes 
$\eta_\psi(s)$; and the relative phase of form factors $\Delta \Phi_\psi(s)$. 
These form-factor combinations are defined in Appendix A.
\begin{figure}[ht]
\begin{center}
 \scalebox{0.6}{\includegraphics{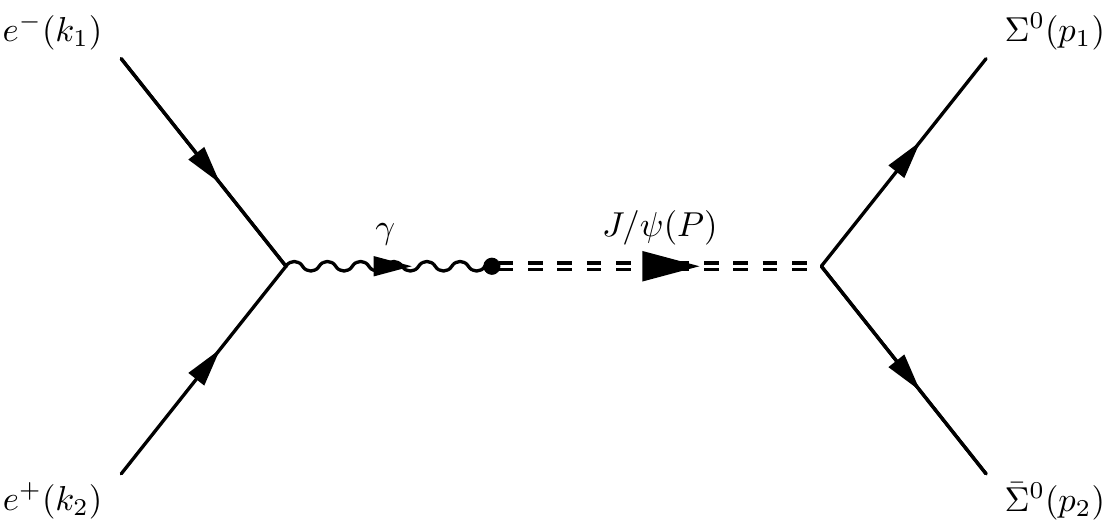}} 
%
\caption{Graph describing the psionic annihilation reaction  
$e^+ e^- \rightarrow J/\psi \rightarrow \bar{\Sigma}^0 \Sigma^0$. The same reaction can also 
proceed hadronicly via  other vector-charmonium states such as 
 $\psi'$ or $\psi(2{\textrm S})$,
 or electromagnetically via photons.}
\label{F1-fig}
\end{center}
\end{figure} 

The theoretical description of the annihilation reaction of Fig.1 can be found   
 in Ref.\cite{GF3}. Accurate experimental results for the form-factor 
parameters $\eta_\psi$ and $\Delta\Phi_\psi$ and the 
weak-interaction parameters $\alpha_\Lambda(\alpha_{\bar{\Lambda}})$  for the $J/\psi$ 
annihilation process are all reported in Ref.\cite{Nature}.
In addition, the graph  can be generalized to include hyperons that decay sequentially.

 Our analysis of the cross-section-distribution function for the 
 annihilation reaction $e^+e^-\rightarrow J/\psi \rightarrow\Sigma^0\bar{\Sigma}^0$, 
 followed by its subsequent hyperon decays, starts from 
the master formula of Ref.\cite{GKa}, and which is reproduced in the following section.
The purpose of our investigation is to find out which coordinate choice  
would be most convenient when 
evaluating the master formula, and at the same time 
being able to compare our result to those of others.
%
%
\section{Master formula}
%
%
In several previous publications we  studied $e^+e^-$ annihilation into hyperon 
pairs $Y\bar{Y}$ and the subsequent decays of those pairs. 
Photon as well as charmonium induced annihilaton was considered.  
In the present investigation we limit ourselves to the  hyperon-decay chain
$\Sigma^0\rightarrow \Lambda \gamma;\Lambda\rightarrow p\pi^- $, and 
its corresponding antihyperon-decay 
chain $\bar{\Sigma}^0\rightarrow\bar{\Lambda} \gamma;\bar{\Lambda}\rightarrow \bar{p}\pi^+ $, again when simultaneously occuring in the reaction $e^+e^-\rightarrow J/\psi\rightarrow\Sigma^0\bar{\Sigma}^0$.

 In  Ref.\cite{GKa} it was shown that the cross-section-distribution function 
 for a $J/\psi$ induced joint production and  subsequent decay 
 of a $\Sigma^0\bar{\Sigma}^0$ pair can be  summarized in the
  {\bfseries master formula}
\begin{equation}
	\rd \sigma =\rd \sigma(e^+e^-\rightarrow J/\psi \rightarrow \Sigma^0\bar{\Sigma}^0)\,
	\Bigg[\frac{\cal{W}({\boldsymbol{\xi}})}{\cal{R}}\Bigg]
	\rd \Phi (\Sigma^0, \Lambda, p; \bar{\Sigma}^0, \bar{\Lambda}, \bar{p}). 
	 \label{MasterForm}
\end{equation}
As can be seen the master formula involves three factors, describing the 
 {\bfseries  annihilation} of a lepton pair into a hyperon pair; 
 the {\bfseries folded product} of spin densities 
${\cal{W}({\boldsymbol{\xi}})}$ representing hyperon production and decay;
  and
 the {\bfseries phase space} element of sequential hyperon decays.
 Each event is specified by a nine-dimensional vector 
${\boldsymbol{\xi}}=(\theta,\Omega_{\Lambda},\Omega_{p},\Omega_{\bar{\Lambda}},\Omega_{\bar{p}}),$
with $\theta$ the scattering angle in the $e^+e^-\rightarrow \Sigma^0\bar{\Sigma}^0$
 subprocess.
 
Following Refs.~\cite{GF3} and \cite{GKa} we write the cross-section-distribution 
function for the $J/\psi$ induced annihilation reaction 
$e^+e^-\rightarrow J/\psi \rightarrow \Sigma^0 \bar{\Sigma}^0$  as
\begin{equation}
	\frac{\rd \sigma}{\rd \Omega_{\Sigma^0}}(e^+e^-\rightarrow J/\psi \rightarrow \Sigma^0 \bar{\Sigma}^0)
	= \frac{p}{k} \frac{\alpha_\psi \alpha_g}{(s-m_\psi^2)^2+m_\psi^2\Gamma(m_\psi)}
	 \,D_\psi(s)
	{\cal{R}} ,\label{Compton-ccc}
\end{equation}
where the strength function $D_\psi(s)$ is defined in Appendix A, and the structure 
function  ${\cal{R}}$ in Appendix B.
 The electromagnetic-coupling constant $\alpha_\psi$  is determined by the  electromagnetic-decay width $\Gamma(J/\psi\rightarrow e^+e^-)$, and 
the hadronic-coupling constant $\alpha_g$ 
 similarly by the  hadronic-decay width $\Gamma(J/\psi\rightarrow \Sigma^0\bar{\Sigma}^0)$.

The differential-spin-distribution function ${\cal{W}}({\boldsymbol{\xi}})$ of Eq.(\ref{MasterForm})
is obtained by {\bfseries folding}  a product of five spin densities,
\begin{equation}
	{\cal{W}}({\boldsymbol{\xi}})={\bigg \langle}S(\mathbf{n}_{\Sigma^0},\mathbf{n}_{\bar{\Sigma}^0})\,
	  G(\mathbf{n}_{\Sigma^0},\mathbf{n}_\Lambda)
	 G(\mathbf{n}_{\Lambda},\mathbf{n}_p)
	 G(\mathbf{n}_{\bar{\Sigma}^0},\mathbf{n}_{\bar{\Lambda}})
	 G(\mathbf{n}_{\bar{\Lambda}},\mathbf{n}_{\bar{p}})
	{\bigg \rangle}_{\mathbf{n}} , \label{Pplus:decayii}
\end{equation}
in accordance with the prescription of Ref.\cite{GF1} and of
Eq.(\ref{Defaverage}).
The folding  operation $\left\langle ...\right\rangle_{\mathbf{n}}$ 
applies to each of the six hadron spin vectors, $\mathbf{n}_{\Sigma^0},...,\mathbf{n}_{\bar{p}}$. 

The function $S(\mathbf{n}_{\Sigma^0}, \mathbf{n}_{\bar{\Sigma}^0})$ represents 
 the spin-density distribution 
for the $\Sigma^0 \bar{\Sigma}^0$ hyperon pair.
 This function also depends on the 
unit vectors $\mathbf{l}_{\Sigma^0}$ and $\mathbf{l}_{\bar{\Sigma}^0}$, 
which are unit vectors in the directions of motion of the 
$\Sigma^0$ and   $\bar{\Sigma}^0$ hyperons in
the center-of-momentum (c.m.) frame of the event. 
The four remaining spin-density-distribution functions 
$G(\mathbf{n}_{Y_1}, \mathbf{n}_{Y_2})$ represent 
spin-density distributions for the hyperon decays $\Sigma^0\rightarrow \Lambda\gamma$; or
$\Lambda\rightarrow p\pi^-$; or their antihyperon counterparts.

The spin-decay-distribution functions 
$G(\mathbf{n}_{Y_1}, \mathbf{n}_{Y_2})$ are normalized to unity,
 which means their spin independent terms are unity. However, for convenience 
the spin-density-distribution function $S(\mathbf{n}_{\Sigma^0},\mathbf{n}_{\bar{\Sigma}^0})$  is normalized to ${\cal{R}}$.

The phase-space factor, $\rd \Phi (\Sigma^0, \Lambda, p; \bar{\Sigma}^0, \bar{\Lambda},\bar{p})$
of the master equation, describes the normalized phase-space element for the
sequential decays of the two baryons $\Sigma^0$ and $\bar{\Sigma}^0$,
\begin{align}
	\rd \Phi (\Sigma^0, \Lambda, p; \bar{\Sigma}^0, \bar{\Lambda},\bar{p})=&  
	\frac{\Gamma(\Sigma^0 \rightarrow \Lambda\gamma)}{\Gamma(\Sigma^0\rightarrow all)}
	\frac{\rd \Omega_{\Lambda}}{4\pi}	\cdot
	\frac{\Gamma(\Lambda \rightarrow p\pi^-)}{\Gamma(\Lambda\rightarrow all)}
	\frac{\rd \Omega_{p}}{4\pi} \nonumber \\ \cdot & 
	\frac{\Gamma(\bar{\Sigma}^0 \rightarrow \bar{\Lambda}\gamma)}
	{\Gamma(\bar{\Sigma}^0\rightarrow all)}
	\frac{\rd \Omega_{\bar{\Lambda}}}{4\pi}\cdot 
	\frac{\Gamma(\bar{\Lambda} \rightarrow \bar{p}\pi^-)}{\Gamma(\bar{\Lambda}\rightarrow all)}
	\frac{\rd \Omega_{\bar{p}}}{4\pi}.
\end{align}
The widths are defined in the usual way. For $\Gamma(\Sigma^0 \rightarrow \Lambda\gamma)$ 
this means forming an average over the $\Sigma^0$ spin directions, and summing 
over the Lambda and gamma 
spin directions.  The angles $\Omega_{\Lambda}$ define the direction of motion of the
$\Lambda$ hyperon in the $\Sigma^0$ rest system; 
the angles $\Omega_{p}$  the direction of motion of the
$p$ baryon in the $\Lambda$ rest system, 
and so on.

%

%
\section{$e^+ e^-$ annihilation into $\Sigma^0 \bar{\Sigma}^0$ pairs}
%
%

The cross-section-distribution function for  $e^+ e^-$ annihilation 
 into a  $\Sigma^0 \bar{\Sigma}^0$ pair appears in two places in the master formula 
 of Eq.(\ref{MasterForm}). The unpolarized-cross-section-distribution function
 is a prefactor in the master formula, and the hyperon-spin-density-distribution function enters as a factor in the 
 spin-density-distribution function of Eq.(\ref{Pplus:decayii}).

The cross-section distribution for polarized-final-state hyperons was derived  
 in Refs.\cite{GKa} and \cite{GF3},
\begin{equation}
	\frac{\rd \sigma}{\rd \Omega_{\Sigma^0}}(e^+e^-\rightarrow J/\psi \rightarrow \Sigma^0 \bar{\Sigma}^0)
	= \frac{p}{4k} \frac{\alpha_\psi \alpha_g D_\psi(s)}  
	             {(s-m_\psi^2)^2+m_\psi^2\Gamma(m_\psi)}
	 \, S(\mathbf{n}_{\Sigma^0},\mathbf{n}_{{\bar{\Sigma}^0}}),
	 \label{Compton-ccc}
\end{equation}
where $D_\psi(s)$ is the strength function of Eq.(\ref{DS_def}),
 $\mathbf{n}_{\Sigma^0}$ and $\mathbf{n}_{\bar{\Sigma}^0}$  the spin vectors of the $\Sigma^0$ and $\bar{\Sigma}^0$ hyperons, and  $S(\mathbf{n}_{\Sigma^0},\mathbf{n}_{{\bar{\Sigma}^0}})$ the spin-density-distribution 
function for the final-state hyperons. This spin-density-distribution function is 
normalized so that its spin-independent part equals $\cal{R}$, with
\begin{equation}
	{ \cal{R}} = 1+\eta_\psi  \cos^2\!\theta, 
\end{equation}
according to Eq.(\ref{DefR}).
Consequently, summing over the final-state-hyperon
polarizations gives the unpolarized cross-section-distribution 
function
\begin{equation}
	\frac{\rd \sigma}{\rd \Omega_{\Sigma^0}}(e^+e^-\rightarrow J/\psi \rightarrow \Sigma^0 \bar{\Sigma}^0)
	= \frac{p}{k} \frac{\alpha_\psi \alpha_g}{(s-m_\psi^2)^2+m_\psi^2\Gamma(m_\psi)}
	 \,D_\psi(s)
	{\cal{R}} ,\label{Compton-cc2c}
\end{equation}

 For a spin-one-half baryon of four-momentum $\mathbf{p}$, the four-vector 
spin $s(p)$ is related to the three-vector-spin direction $\mathbf{n}$, the spin in 
the rest system,  by
\begin{equation}	s(\mathbf{p},\mathbf{n})=\frac{n_\|}{M}(|\mathbf{p}|,E\hat{\mathbf{p}})+
(0,\mathbf{n}_\bot ).
	\label{4spin}
\end{equation}
Longitudinal and transverse directions of vectors are relative 
to the $\hat{\mathbf{p}}$ direction.

In the global c.m.\ system kinematics simplifies. 
There, three-momenta $\mathbf{p}$ and $\mathbf{k}$ are defined such that
	\begin{align}
	\mathbf{p}_{\Sigma^0}  &=  - \mathbf{p}_{\bar{\Sigma}^0}= \mathbf{p} , \\
	\mathbf{k}_{e^+} &= - \mathbf{k}_{e^-} = \mathbf{k}, 
\end{align}
	and the scattering angle $\theta$ such that
	$\cos\theta= \hat{\mathbf{p}}\cdot \hat{\mathbf{k}}.$ 
	 For the $\Sigma^0$ and $\bar{\Sigma}^0$   
unit vectors $\mathbf{l}_{\Sigma^0}$ and $\mathbf{l}_{\bar{\Sigma}^0}$, 
we have 
$\mathbf{l}_{\Sigma^0}=-\mathbf{l}_{\bar{\Sigma}^0}=\hat{\mathbf{p}}$.

The spin-density-distribution function 
 $S(\mathbf{n}_{\Sigma^0},\mathbf{n}_{2})$ is a sum of
 seven mutually orthogonal contributions \cite{GF2},
\begin{align}
	S(\mathbf{n}_{\Sigma^0},\mathbf{n}_{\bar{\Sigma}^0}) =& \,  {\cal{R}}  
	    + {\cal{S}}\, \mathbf{N}\cdot \mathbf{n}_{\Sigma^0}
	+ {\cal{S}}\,\mathbf{N}\cdot \mathbf{n}_{\bar{\Sigma}^0}
	 + {\cal{T}}_{1}
	  \mathbf{n}_{\Sigma^0} \cdot \hat{\mathbf{p}}\mathbf{n}_{\bar{\Sigma}^0} \cdot \hat{\mathbf{p}}
	    \nonumber\\   
	&  + {\cal{T}}_2  \mathbf{n}_{{\Sigma^0}\bot} \cdot \mathbf{n}_{{\bar{\Sigma}^0}\bot }
		+ {\cal{T}}_3 \mathbf{n}_{{\Sigma^0}\bot} \cdot \hat{\mathbf{k}}\mathbf{n}_{{\bar{\Sigma}^0}\bot}
		\cdot \hat{\mathbf{k}}/  \sin^ 2    \!\theta
		\nonumber\\
			&  + {\cal{T}}_4  \bigg( \mathbf{n}_{\Sigma^0} \cdot \hat{\mathbf{p}}\mathbf{n}_{{\bar{\Sigma}^0}\bot} \cdot \hat{\mathbf{k}} 		
			 + \mathbf{n}_{\bar{\Sigma}^0} \cdot \hat{\mathbf{p}}\mathbf{n}_{{\Sigma^0}\bot} \cdot \hat{\mathbf{k}} 
			  \bigg)/\sin \theta  ,\label{ISD}
\end{align}
where $\mathbf{N}$ is normal to the scattering plane,

\begin{equation}
 \mathbf{N}=\frac{1}{\sin\theta} \,  \hat{\mathbf{p}}\times \hat{\mathbf{k}}.\label{Ndef}
\end{equation}
The six structure functions ${\cal{R}}$, ${\cal{S}}$, and ${\cal{T}}$ of Eq.(\ref{ISD})  
 depend on the scattering angle $\theta$, the ratio function $\eta_\psi(s)$, and
the phase function $\Delta\Phi_\psi(s)$. 
For their definitions we refer to Appendix \ref{AppC}, but be careful, 
 our original definitions 
were slightly different \cite{GF2}.
 
%
%
%
\section{Assorted spin densities}\label{fyra}
%

To be able to calculate the differential-distribution function of 
Eq.(\ref{Pplus:decayii}) we need in addition to the 
spin-density-distribution function 
for the $\Sigma^0\bar{\Sigma}^0$ final-state  pair, 
the spin-density-distribution functions for the decays $\Sigma^0\rightarrow\Lambda\gamma $  
and $\Lambda\rightarrow p\pi^-$, and their antiparticle conjugate decays.

Weak decays of spin-one-half baryons, such as $\Lambda\rightarrow p\pi^-$,
 involve  both S-
and P-wave amplitudes, and the spin-density-decay distribution is commonly parametrized by  
three  parameters, denoted $\alpha\beta\gamma$,  and which fulfill a relation 
\begin{equation}
	\alpha^2 + \beta^2 +\gamma^2=1.
\end{equation}
Details of this description can be found in Refs.\cite{Lee} and \cite{GKa}. 

The spin-density-distribution function 
$G(\mathbf{n}_{\Lambda},\mathbf{n}_{p} )$, 
describing the decay $\Lambda\rightarrow p\pi^-$, is a scalar, which we choose 
to evaluate in the rest system of the $\Lambda$ hyperon, to get
\begin{equation} 	
		G(\mathbf{n}_{\Lambda},\mathbf{n}_{p} ) = 1+ \alpha_{\Lambda} \mathbf{n}_{\Lambda}\cdot \mathbf{l}_{p}
	  +\alpha_{\Lambda} \mathbf{n}_{p}\cdot \mathbf{l}_{p}
	 +\mathbf{n}_{\Lambda}\cdot 
	\mathbf{L}_{\Lambda}(\mathbf{n}_{p},\mathbf{l}_{p} ),\label{weakG}
\end{equation}	
with the vector-valued function
 $\mathbf{L}_\Lambda (\mathbf{n}_p, \mathbf{l}_p)$ defined as
\begin{equation}
\mathbf{L}_\Lambda (\mathbf{n}_p, \mathbf{l}_p)
	=\,\gamma_\Lambda  \mathbf{n}_p
	+\big[(1-\gamma_\Lambda )\mathbf{n}_p\cdot \mathbf{l}_p\big]\, \mathbf{l}_p
	+\beta_\Lambda   \mathbf{n}_p\times \mathbf{l}_p .\label{Gccd_2}
\end{equation} 
Here, $\mathbf{n}_{\Lambda}$ and $\mathbf{n}_p$ are
 the spin vectors of the $\Lambda$ hyperon and the $p$ baryon, and 
 $\mathbf{l}_p$ a unit vector in the direction of motion of the
proton in the rest system of the $\Lambda$ hyperon. The $\Lambda$ indices 
remind us  the parameters refer to a $\Lambda$ decay.
An important aspect of the spin-density-distribution function 
is its normalization. The spin-independent term is unity.

The spin-density-distribution function 
$G(\mathbf{n}_{\bar{\Lambda}},\mathbf{n}_{\bar{p}} )$
 for the antiparticle-conjugate decay $\bar{\Lambda }\rightarrow \bar{p} \pi^+$  has exactly the same functional structure as $G(\mathbf{n}_{\Lambda},\mathbf{n}_{p} )$, 
 but the decay parameters take other numerical values. 
For  {\slshape{CP}} conserving interactions the asymmetry parameters  
of the $\Lambda$-hyperon decay
are related to those of the $\bar{\Lambda}$-hyperon decay  by \cite{Don1,Don2}
\begin{equation}
	\alpha_\Lambda =-\alpha_{\bar{\Lambda }},\quad \beta_\Lambda =-\beta_{\bar{\Lambda }}, \quad 
	   \gamma_\Lambda =\gamma_{\bar{\Lambda }}.
\end{equation}

Next, we turn to the electromagnetic M1 transition 
$\Sigma^0\rightarrow \Lambda\gamma$. It is caused by a 
transition-magnetic moment, of strength
\begin{equation}
	\mu_{\Sigma \Lambda}=eF_2(0)/(m_\Sigma +m_\Lambda ).
\end{equation}

The normalized-spin-density-distribution function for a $\Sigma^0\rightarrow \Lambda\gamma$ transition 
to a final state of fixed photon helicity $\lambda_\gamma$ is, 
according to Ref.\cite{GKa}, 
\begin{equation}
	G_\gamma(\mathbf{n}_{\Sigma^0},\mathbf{n}_{\Lambda};\lambda_\gamma)=
	1-
	\mathbf{n}_{\Sigma^0}\cdot \mathbf{l}_{\gamma}\,  \mathbf{l}_{\gamma}\cdot \mathbf{n}_{\Lambda }
	+ \lambda_\gamma (\mathbf{n}_{\Sigma^0}\cdot \mathbf{l}_{\gamma}
	-\mathbf{n}_{\Lambda}\cdot \mathbf{l}_{\gamma} ),
	\label{Photav}
\end{equation}	
where $\mathbf{l}_{\gamma}$ is a unit vector in the direction of motion of the photon, 
and $ \mathbf{l}_{\Lambda }=-\mathbf{l}_{\gamma}$ a unit vector in the direction of motion of
the $\Lambda$ hyperon, both in the rest system of the $\Sigma^0 $ baryon.
The photon  helicities $\lambda_\gamma$ take on the values $\pm 1$.
 
We notice that when both hadron spins are parallel or anti-parallel to the photon
momentum, then the decay probability vanishes, a property of angular-momentum conservation.

Summing, in Eq.(\ref{Photav}),  the contributions from 
the two photon-helicity states gives the 
normalized-spin-density-distribution function 
\begin{equation}
	G(\mathbf{n}_{\Sigma^0},\mathbf{n}_{\Lambda})= 1-
	\mathbf{n}_{\Sigma^0}\cdot \mathbf{l}_{\gamma}\,  \mathbf{l}_{\gamma}\cdot \mathbf{n}_{\Lambda}.
	\label{Photav}
\end{equation}	

The normalized-spin-density-distribution function 
for the conjugate transition, 
$\bar{\Sigma}^0\rightarrow \bar{\Lambda}\gamma$, 
is obtained by replacing, in expression (\ref{Photav}), the particle 
spin vectors $\mathbf{n}_{\Sigma^0}$ and $\mathbf{n}_{\Lambda} $
by the antiparticle-spin vectors $\mathbf{n}_{\bar{\Sigma}^0}$
 and $\mathbf{n}_{\bar{\Lambda}}$ .

%
\section{Sequential decay of hyperons}
%

A factor of our master formula for hyperon production and decay, 
Eq.(\ref{MasterForm}), is
the differential-spin-distribution function 
${\cal{W}}({\boldsymbol{\xi}})$ of Eq.(\ref{Pplus:decayii}),
which is obtained by  folding  a product of five spin densities. 
The folding prescription  is especially adapted to spin one-half baryons. 
A folding operation  implies forming an average over
 intermediate-spin directions $\mathbf{n}$ according to the prescription
 of Refs.\cite{GF1}
 and \cite{GFjuni}
\begin{equation}
\big{\langle } 1\big{\rangle }_{\mathbf{n}}   =1, \quad 
\big{\langle }  \mathbf{n} \big{\rangle }_{\mathbf{n}}   =0, \quad
\big{\langle }  \mathbf{n}\cdot \mathbf{k}  \mathbf{n}\cdot \mathbf{l}  \big{\rangle }_{\mathbf{n}}   
=\mathbf{k}\cdot \mathbf{l} .\label{Defaverage}
 \end{equation}
\begin{figure}[h]
\begin{center}
%
\centerline{\scalebox{0.50}{ \includegraphics{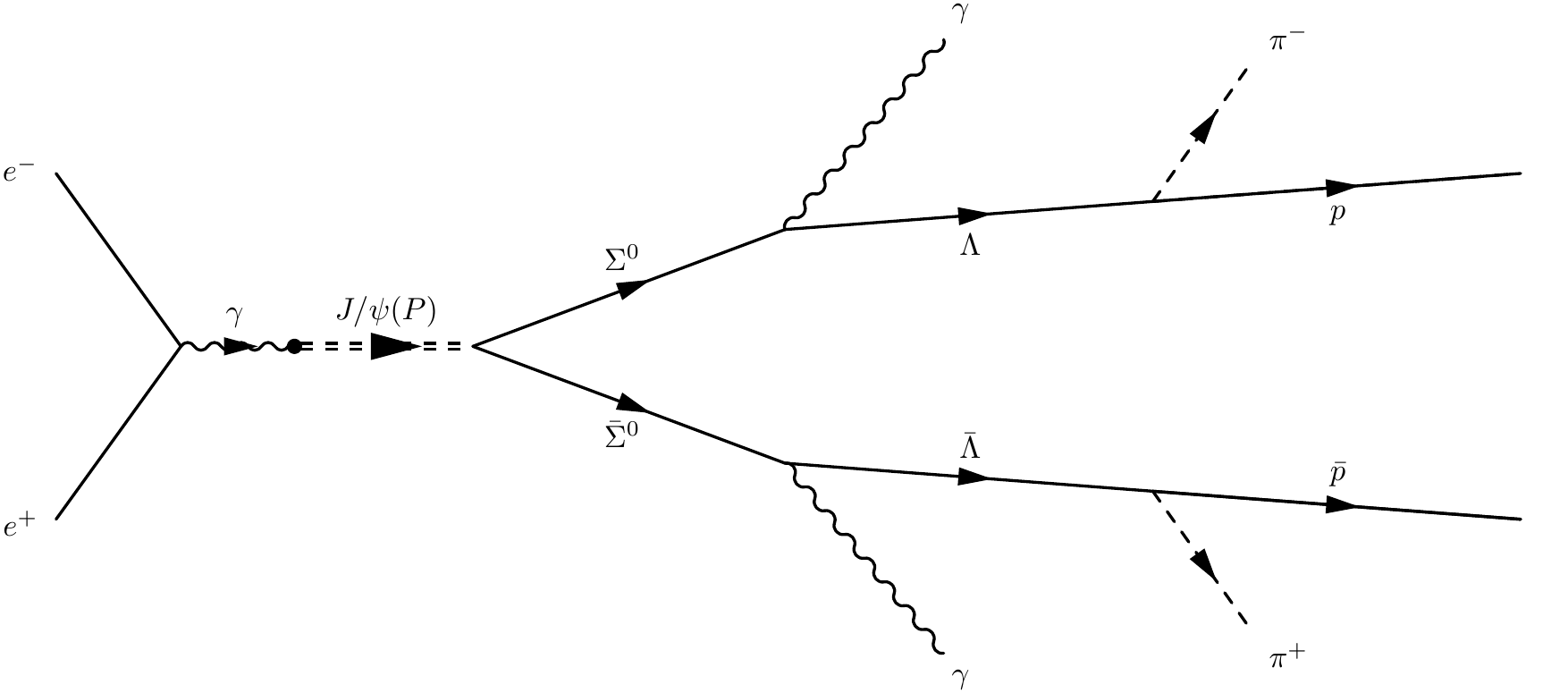} }} 
      \caption{Graph describing the reaction 
$e^+ e^- \rightarrow  \bar{\Sigma}^0 \Sigma^0$, and the subsequent decays, 
$\Sigma^0\rightarrow\Lambda\gamma;\Lambda\rightarrow p\pi^-$  
and $\bar{\Sigma}^0\rightarrow\bar{\Lambda}\gamma;\bar{\Lambda} \rightarrow \bar{p}\pi^+$.
 The reaction graphed can, in addition to photons, 
be mediated by   vector charmonia, such as  $J/\psi$, 
$\psi'$ and $\psi(2{\textrm S})$. Solid lines refer to baryons, dashed to mesons, and wavy to photons.}
\end{center}
\label{F200-fig}
\end{figure}
The  spin-density distribution
 $W(\mathbf{n}_{{\Sigma^0}},\mathbf{n}_p)$ for the  decay chain 
$\Sigma^0\rightarrow \Lambda \gamma; \Lambda\rightarrow p\pi^-$ is 
obtained by folding the product of the spin 
density distributions in the decay chain.
We obtain
\begin{equation}
	W(\mathbf{n}_{\Sigma^0},\mathbf{n}_p) ={\bigg\langle}
	G(\mathbf{n}_{\Sigma^0},\mathbf{n}_\Lambda)
	G(\mathbf{n}_\Lambda,\mathbf{n}_p){\bigg\rangle}_{\mathbf{n}_{\Lambda}},
\end{equation}
where the two spin-density-distribution functions 
on the right-hand side are defined in Eqs.(\ref{Photav}) and (\ref{weakG}).
Performing the folding operation gives 
\begin{align}
	W(\mathbf{n}_{\Sigma^0},\mathbf{n}_p) &= U_{\Sigma^0} +  \mathbf{n}_{\Sigma^0}\cdot 
	\mathbf{V}_{\Sigma^0}, \label{Yabarchain}\\
	U_{\Sigma^0} &=       1+  \alpha_{\Lambda} \mathbf{n}_{p}\cdot \mathbf{l}_{p}   ,     \label{Yachain}       \\
	\mathbf{V}_{\Sigma^0} &=- \mathbf{l}_{\gamma}  \left[ 
	  \alpha_{\Lambda} \mathbf{l}_{\gamma}\cdot \mathbf{l}_p 
	          + \mathbf{n}_{p}\cdot {\mathbf{L}}_{\Lambda}(\mathbf{l}_{\gamma}, -\mathbf{l}_{p})
	  \right],
			\label{Ybchain}
\end{align}
and ditto for $W(\mathbf{n}_{\bar{\Sigma}^0},\mathbf{n}_{\bar{p}})$.

%
%
\section{Production and  decay of $\Sigma^0 \bar{\Sigma}^0 $ pairs}
%
%

Now, we come to our final task; 
production and  decay of $\Sigma^0 \bar{\Sigma}^0 $ pairs.
 The starting point 
is the reaction $e^+e^-\rightarrow \Sigma^0 \bar{\Sigma}^0$, 
the spin-density-distribution function of which was calculated in Sect.3, 
and named $S(\mathbf{n}_{\Sigma^0},\mathbf{n}_{\bar{\Sigma}^0})$. 
The  spin-density-distribution function 
 $W(\mathbf{n}_{{\Sigma}^0},\mathbf{n}_p)$ which represents 
the  decay chain 
$\Sigma^0\rightarrow \Lambda \gamma; \Lambda\rightarrow p\pi^-$ was 
calculated in Sect.~5; 
and so for the anti-chain-decay function 
$W(\mathbf{n}_{\bar{\Sigma}^0},
\mathbf{n}_{\bar{p}})$.

The final-state-angular distributions are obtained by folding the
spin distributions for production and decay, according to presciption (\ref{Defaverage}). 
Invoking Eq.(\ref{ISD}) for the production step and 
 Eq.(\ref{Yabarchain}) and its anti-distribution for the  decay steps, we get the  differential-spin-density-distribution function 
\begin{align}
	{\cal{W}}({\boldsymbol{\xi}})=& \,{\bigg\langle}
	     S(\mathbf{n}_{\Sigma^0}, \mathbf{n}_{\bar{\Sigma}^0})   
	 	W(\mathbf{n}_{\Sigma^0},\mathbf{n}_p)W
		(\mathbf{n}_{\bar{\Sigma}^0},\mathbf{n}_{\bar{p}})
	 	{\bigg\rangle}_{\mathbf{n}_{\Sigma^0}, \mathbf{n}_{\bar{\Sigma}^0}} 
		\nonumber \\
	 =& \,  {\cal{R}} U_{\Sigma^0} U_{\bar{\Sigma}^0 }
	    + {\cal{S}}U_{\bar{\Sigma}^0 }\, \mathbf{N}\cdot \mathbf{V}_{\Sigma^0}
	+ {\cal{S}}U_{\Sigma^0}\,\mathbf{N}\cdot \mathbf{V}_{\bar{\Sigma}^0} 
	  \nonumber \\
	& + {\cal{T}}_1 
	  \mathbf{V}_{\Sigma^0} \cdot \hat{\mathbf{p}}\mathbf{V}_{\bar{\Sigma}^0} \cdot \hat{\mathbf{p}}
	    + {\cal{T}}_2  \mathbf{V}_{\Sigma^0\bot} \cdot \mathbf{V}_{\bar{\Sigma}^0\bot} \nonumber \\ &
		+ {\cal{T}}_3 \mathbf{V}_{\Sigma^0\bot} \cdot \hat{\mathbf{k}}\mathbf{V}_{\bar{\Sigma}^0\bot}
		\cdot \hat{\mathbf{k}}/\sin^2\! \theta
		\nonumber\\
			&  + {\cal{T}}_4  \bigg( \mathbf{V}_{\Sigma^0} \cdot \hat{\mathbf{p}}
			\mathbf{V}_{\bar{\Sigma}^0\bot} \cdot \hat{\mathbf{k}} 
			 + \mathbf{V}_{\bar{\Sigma}^0} \cdot 
			\hat{\mathbf{p}}\mathbf{V}_{\Sigma^0\bot} \cdot \hat{\mathbf{k}} 
			  \bigg)/\sin \theta  .\label{Hdef2}
\end{align}
The functions $U_{\Sigma^0} $ and $\mathbf{V}_{\Sigma^0}$
 are defined in Sect.~5, and
\begin{align}
	U_{\Sigma^0} &=   1+  \alpha_{\Lambda} \mathbf{n}_{p}\cdot \mathbf{l}_{p} ,     \label{Yachain}       \\
	\mathbf{V}_{\Sigma^0} &=- \mathbf{l}_{\gamma}  \left[ 
	  \alpha_{\Lambda} \mathbf{l}_{\gamma}\cdot \mathbf{l}_p 
	          + \mathbf{n}_{p}\cdot {\mathbf{L}}_{\Lambda}(\mathbf{l}_{\gamma}, -\mathbf{l}_{p})
	  \right].
			\label{Ybchain}
\end{align}
We observe that $U_{\Sigma^0} $ depends on 
the weak interaction parameter $\alpha_{\Lambda} $, 
whereas $\mathbf{V}_{\Sigma^0}$ in addition depends on the parameters
 $\beta_{\Lambda} $ and $\gamma_{\Lambda} $ 
through the vector function ${\mathbf{L}_\Lambda}$, of Eq.(\ref{Gccd_2}).

The angular distributions of Eq.(\ref{Hdef2}) are the most 
general ones, and still depend on the spin vectors
 $\mathbf{n}_p$ and $\mathbf{n}_{\bar{p}}$ which are difficult to measure. If we 
are willing to consider proton- and anti-proton-spin averages, 
then variables $U$ and $\mathbf{V}$ simplify,
\begin{align}
U_{\bar{\Sigma}^0}&=1 , \qquad  \mathbf{V}_{{\Sigma}^0}=
		-\alpha_{{\Lambda}}\, \mathbf{l}_{{\Lambda}}\cdot 
		 \mathbf{l}_{{p}}\mathbf{l}_{{\Lambda} }, \nonumber\\
		U_{\bar{\Sigma}^0}&=1 , \qquad  \mathbf{V}_{\bar{\Sigma}^0}=
		-\alpha_{\bar{\Lambda}}\, \mathbf{l}_{\bar{\Lambda}}\cdot 
		 \mathbf{l}_{\bar{p}}\mathbf{l}_{\bar{\Lambda} } . \label{UVSig}
\end{align}
Since $U_{\Sigma^0}=U_{\bar{\Sigma}^0}=1$ the effect of the folding is to make, in 
the spin-density function $S(\mathbf{n}_{\Sigma^0},\mathbf{n}_{\bar{\Sigma}^0})$ of Eq.(\ref{ISD}), 
 the replacements 
$\mathbf{n}_{\Sigma^0}\rightarrow  \mathbf{V}_{\Sigma^0}$ and 
$\mathbf{n}_{\bar{\Sigma}^0}\rightarrow  \mathbf{V}_{\bar{\Sigma}^0}$.
We notice that the $U$ and $\mathbf{V}$ variables are independent of the
weak-asymmetry parameters $\beta_\Lambda$ and $\gamma_\Lambda$. 
Their dependence is hidden in the vector function 
${\mathbf{L}}_{\Lambda}(\mathbf{l}_{\gamma}, -\mathbf{l}_{p})$ of 
Eq.(\ref{Yabarchain}), and which is absent in Eq.(\ref{Hdef2}).

Inserting the expressions of Eq.(\ref{UVSig}) into the spin-density function 
of Eq.(\ref{Hdef2}), we get
\begin{align}
	 {\cal{W}}({\boldsymbol{\xi}})
	 =& \,  {\cal{R}} 
	    -\alpha_{\Lambda} {\cal{S}}\, \mathbf{N}\cdot \mathbf{l}_{\Lambda}\,  \mathbf{l}_{\Lambda} \cdot \mathbf{l}_p
	-\alpha_{\bar{\Lambda}} {\cal{S}}\,\mathbf{N}\cdot 
	\mathbf{l}_{\bar{\Lambda}}\,  
	           \mathbf{l}_{\bar{\Lambda}} \cdot \mathbf{l}_{\bar{p}}
	 \nonumber \\
	& + \alpha_{\Lambda}\alpha_{\bar{\Lambda}}\,   \mathbf{l}_{\Lambda}\cdot \mathbf{l}_p
	  \mathbf{l}_{\bar{\Lambda}} \cdot \mathbf{l}_{\bar{p}}
	\, \bigg{[}\,  {\cal{T}}_1                          
	   \mathbf{l}_\Lambda \cdot \hat{\mathbf{p}}
		                 \mathbf{l}_{\bar{\Lambda}} \cdot \hat{\mathbf{p}} \nonumber\\
	   & + {\cal{T}}_2  \mathbf{l}_{{\Lambda}\bot} \cdot \mathbf{l}_{\bar{\Lambda}\bot} 
	 + {\cal{T}}_3 \mathbf{l}_{\Lambda\bot} \cdot \hat{\mathbf{k}}\mathbf{l}_{\bar{\Lambda}\bot}
	\cdot \hat{\mathbf{k}}/\sin^2\! \theta
		\nonumber\\
			&  + {\cal{T}}_4  \bigg( \mathbf{l}_{\Lambda} \cdot \hat{\mathbf{p}} 
			\mathbf{l}_{\bar{\Lambda}\bot} \cdot \hat{\mathbf{k}} 
			 + \mathbf{l}_{\bar{\Lambda}} \cdot \hat{\mathbf{p}}\mathbf{l}_{\Lambda\bot} \cdot \hat{\mathbf{k}} 
			  \bigg)/\sin \theta \bigg{]} .\label{Hdef3a}
\end{align}
Thus, this is the angular distribution obtained when  folding the product of spin densities 
for production and decay. These results were previously reported in 
Refs.\cite{GKa} and \cite{GFjuni}.
%
%
%
%
%
\section{Differential-spin distributions}\label{Sec.Dsb}
%

A closer inspection of the differential-spin-density-distribution function
of Eq.(\ref{Hdef3a}) shows that the weak-interaction
parameters $\alpha_\Lambda$ and $\alpha_{\bar{\Lambda}}$ 
always come in the combinations
$\alpha_\Lambda\mathbf{l}_{{\Lambda}}\cdot \mathbf{l}_{{p}}$
or  $\alpha_{\bar{\Lambda}}\mathbf{l}_{\bar{{\Lambda}}}
  \cdot \mathbf{l}_{\bar{{p}}}$. 
	Therefore, it is convenient to define   the following functions;
\begin{align}
\lambda_\Lambda(\theta_{\Lambda p}) &= \alpha_\Lambda 
  \mathbf{l}_{{\Lambda}}\cdot \mathbf{l}_{{p}}
 =\alpha_\Lambda \cos(\theta_{\Lambda p}), \label{Ang1} \\
	\lambda_{\bar{\Lambda}}(\theta_{\bar{\Lambda}\bar{p}}) &= 
	\alpha_{\bar{\Lambda}}
	 \mathbf{l}_{\bar{\Lambda}}\cdot \mathbf{l}_{\bar{p}} = 
          \alpha_{\bar{\Lambda} } \cos(\theta_{\bar{\Lambda}\bar{p}}).
					\label{Ang2}
\end{align}
Then, the    differential-spin-density-distribution function 
of Eq.(\ref{Hdef3a}) can be rewritten as
\begin{equation}
\begin{split}
{\cal{W}}({\boldsymbol{\xi}})=&\ {\cal{R}}- 
		\bigg[ \lambda_\Lambda{{Q}}_\Lambda + 
		\lambda_{\bar{\Lambda}}Q_{\bar{\Lambda}}
		\bigg] {\cal{S}}       \\&             
   +\lambda_\Lambda\lambda_{\bar{\Lambda}}\bigg[  Q_1 {\cal{T}}_1 +
    Q_2 {\cal{T}}_2 + Q_3 {\cal{T}}_3+ Q_4 {\cal{T}}_4 \bigg]
    , \label{Weqn:pdf}
\end{split}
\end{equation}
with the argument ${\boldsymbol{\xi}}$  a nine-dimensional vector
 ${\boldsymbol{\xi}}=(\theta,\Omega_{\Lambda},\Omega_{p},\Omega_{\bar{\Lambda}},\Omega_{\bar{p}})$ representing the scattering angle and 
four directional-unit vectors of particle motion.  

The six structure 
functions ${\cal{R}}$, ${\cal{S}}$, and ${\cal{T}}$ 
are functions of the scattering angle $\theta$ and the 
ratio of form factors $\eta_\psi$.
The six kinematic $Q$ functions are functions of 
$ \mathbf{l}_{\Lambda}$ and $ \mathbf{l}_{\bar{\Lambda}}$.
Their dependencies on the unit vectors
 $\mathbf{l}_{{p}}$ and $\mathbf{l}_{{\bar{p}}}$
reside solely in the functions $\lambda_\Lambda$ and $\lambda_{\bar{\Lambda}}$
of Eqs.(\ref{Ang1}) and (\ref{Ang2}).

The analytic expressions for the six  functions
$Q(\mathbf{l}_{\Lambda},\mathbf{l}_{\bar{\Lambda}})$ are obtained 
by comparing Eqs.(\ref{Hdef3a}) and (\ref{Weqn:pdf});
\begin{align}     
	Q_{\Lambda} =& \mathbf{N}\cdot \mathbf{l}_\Lambda,  \nonumber\\
	Q_{\bar{\Lambda}}  =& \mathbf{N}\cdot \mathbf{l}_{\bar{\Lambda}} 	,\nonumber\\
	 Q_1 =&\mathbf{l}_{\Lambda}\cdot \hat{\mathbf{p}}
                  \mathbf{l}_{\bar{\Lambda}}\cdot \hat{\mathbf{p}},  \nonumber\\
	Q_2 =&\mathbf{l}_{\Lambda\bot} \cdot 
	          \mathbf{l}_{\bar{\Lambda}\bot} ,\nonumber\\
	Q_3=&   
	\mathbf{l}_{\Lambda\bot} \cdot \hat{\mathbf{k}}
		          \mathbf{l}_{\bar{\Lambda}\bot} \cdot \hat{\mathbf{k}}
							/\sin^2\! \theta
								 ,\nonumber\\
  Q_4 =& \bigg[ \mathbf{l}_{\Lambda}\cdot \hat{\mathbf{p}}
\mathbf{l}_{\bar{\Lambda}\bot} \cdot \hat{\mathbf{k}} +
			\mathbf{l}_{\bar{\Lambda}} \cdot \hat{\mathbf{p}}
			\mathbf{l}_{{\Lambda} \bot}\cdot\hat{\mathbf{k}} 
			\bigg] /\sin\theta.
			\label{SixQ}
\end{align}  
Here, longitudinal  and transverse components of vectors 
are defined relative 
 to  $\hat{\mathbf{p}}$, the direction of motion of the 
$\Sigma^0$ hyperon.

The differential-spin-density  distribution of Eq.(\ref{Weqn:pdf}), and the angular functions above, 
 depend on a number of unit vectors; 
$\hat{\mathbf{p}}$ and $-\hat{\mathbf{p}}$ are unit vectors along the directions of motion of the $\Sigma^0$ 
and the $\bar{\Sigma}^0$  in the c.m. system;  
$\hat{\mathbf{k}}$  and $-\hat{\mathbf{k}}$ are unit vectors along the directions of motion of the incident electron 
and positron in the c.m. system;
$\mathbf{l}_{\Lambda}$ and $\mathbf{l}_{\bar{\Lambda}}$ are unit vectors along the 
directions of motion of the $\Lambda$ and $\bar{\Lambda}$ in the rest systems of the $\Sigma^0$ 
and the $\bar{\Sigma}^0$; 
$\mathbf{l}_{p}$ and $\mathbf{l}_{\bar{p}}$ are unit vectors along the 
directions of motion of the $p$ and the $\bar{p}$ in the rest systems of the $\Lambda$ 
and the $\bar{\Lambda}$. 
%
%
\section{Global angular functions}

%
%
The differential-spin-density distribution (\ref{Hdef3a}) is a function of several unit vectors. In order to handle them we need a common coordinate system, which we call global and define as follows.
The scattering plane  of the reaction 
$e^+e^-\rightarrow\Sigma^0\bar{\Sigma}^0$ is spanned by the unit vectors 
 $\hat{\mathbf{p}}=\mathbf{l}_{\Sigma^0}$ and 
$\hat{\mathbf{k}}=\mathbf{l}_{e}$, as 
measured in  the c.m.\ system. 
 The scattering plane makes up 
the $xz$-plane, with the $y$-axis   along the normal to the scattering plane. 
We choose a right-handed coordinate system with basis vectors 
	\begin{eqnarray}
	\mathbf{e}_z  &=&  \hat{\mathbf{p}}, \nonumber\\
	\mathbf{e}_y  &=& \frac{1}{\sin\theta } ( \hat{\mathbf{p}}\times \hat{\mathbf{k}} ) ,
	\nonumber \\
	\mathbf{e}_x  &=& \frac{1}{\sin\theta } (\hat{\mathbf{p}}\times \hat{\mathbf{k}} ) 
	 \times\hat{\mathbf{p}},\label{xaunity}
\end{eqnarray}
and where the initial-state-lepton momentum is decomposed as
\begin{equation}
	\hat{\mathbf{k}}= \sin\theta\,  \mathbf{e}_x +\cos\theta\,	\mathbf{e}_z  .
\end{equation}
The reason we call this coordinate system global is that we use it whenever  studying a sub-process of the $e^+e^-$ annihilation.

In spherical $xyz$ coordinates the unit vectors $\mathbf{l}_{\Lambda}$ 
and $\mathbf{l}_{\bar{\Lambda}}$
associated with  the directions of motion of 
the $\Lambda$ and $\bar{\Lambda}$ hyperons   are,
\begin{eqnarray}
	\mathbf{l}_\Lambda &=&(\cos \phi_{\Lambda} \sin \theta_{\Lambda},  
	\sin \phi_\Lambda \sin \theta_\Lambda, \cos \theta_\Lambda),\nonumber\\
\mathbf{l}_{\bar{\Lambda}}&=&(\cos \phi_{\bar{\Lambda}}
	\sin \theta_{\bar{\Lambda}},  \sin \phi_{\bar{\Lambda}}
	\sin \theta_{\bar{\Lambda}}, \cos \theta_{\bar{\Lambda}}).
	\label{E-def}
\end{eqnarray}
However, 
in order to make our formulas more transparent we introduce 
the notations $\mathbf{l}_{\Lambda} =\mathbf{E}=(E_x, E_y, E_z)$ 
and $\mathbf{l}_{\bar{\Lambda}} =\mathbf{F}=(F_x, F_y, F_z).$ 
In this Cartesian notation, the expressions for kinematic functions
 $Q(\mathbf{l}_{\Lambda},\mathbf{l}_{\bar{\Lambda}} )$ 
of Eq.(\ref{SixQ}) are,
\begin{align}     
	Q_{\Lambda} &= E_y, \qquad &
	Q_{\bar{\Lambda}}  &= F_y ,  \nonumber\\
	 Q_1& = E_z F_z,  \qquad &
	Q_2 &= E_x F_x+E_y F_y ,\nonumber\\
	Q_3 &= E_x F_x,	&	\qquad				 
  Q_4 &= E_xF_z + E_z F_x . \label{SixEE}
\end{align} 
Inserting these expressions for the $Q$ functions into Eq.(\ref{Weqn:pdf}), the definition of the
differential-spin-density-distribution function, gives
\begin{equation}
\begin{split}
{\cal{W}}({\boldsymbol{\xi}}(\Omega))=&\ 1+\eta_\psi\cos^2\! \theta 
     \\ 
		& -\sqrt{1-\eta_\psi^2}\sin(\Delta \Phi_\psi)\sin\theta \cos\theta
		\bigg[\lambda_{\Lambda}E_y+ \lambda_{\bar{\Lambda}} F_y\bigg]      \\
		&    +\lambda_{\Lambda} \lambda_{\bar{\Lambda}} 
	\bigg[ (1+\eta_\psi ) E_z F_z
	+\sin^2\! \theta \big( E_xF_x -E_zF_z 
	  -\eta_\psi E_y F_y\big)  
	\\ &+
	\  \sqrt{1-{{\eta_\psi}}^2}\cos({{\Delta\Phi_\psi}})\sin\theta \cos\theta\,  
	    (E_x F_z + E_z F_x ) \bigg]. \label{eqn:GK}
\end{split}
\end{equation}
The phase-space-angular variables are hidden inside 
the $\mathbf{E}(\theta_\Lambda, \phi_\Lambda)$ and 
$\mathbf{F}(\theta_{\bar{\Lambda}}, \phi_{\bar{\Lambda}})$ functions.

The differential-spin-density-distribution function 
${\cal{W}}({\boldsymbol{\xi}})$ of Eq.(\ref{eqn:GK}) involves two parameters related to the $e^+e^-\to\Sigma^0\bar{\Sigma}^0$ 
reaction that can be determined by data: the ratio of form factors 
$\eta_{\psi}$,  and the relative phase of form factors $\Delta\Phi_{\psi}$. 
In addition, the distribution function ${\cal{W}}({\boldsymbol{\xi}})$ depends on the
 weak-asymmetry parameters 
$\alpha_\Lambda$ and $\alpha_{\bar{\Lambda}}$ of the two Lambda-hyperon decays.
The dependencies on the weak-asymmetry parameters $\beta$ and $\gamma$ drop out, when final-state-proton 
and antiproton spins are not measured.

An important conclusion to be drawn from the differential distribution of Eq.(\ref{eqn:GK}) is 
that when the phase $\Delta \Phi_{\psi}$ is small, the parameters $\alpha_\Lambda$ and $\alpha_{\bar{\Lambda}}$
 are strongly correlated and therefore difficult 
to separate. In order to contribute to the experimental precision value of 
$\alpha_\Lambda$ and $\alpha_{\bar{\Lambda}}$ a non-zero value
 of $\Delta \Phi_{\psi}$ is required.

%
%
%
\section{Helicity angular functions}
%

In the helicity-coordinate system, 
the scattering plane  of the reaction 
$e^+e^-\rightarrow\Sigma^0\bar{\Sigma}^0$ is still spanned by the unit vectors 
 $\hat{\mathbf{p}}=\mathbf{l}_{\Sigma^0}$ 
and $\hat{\mathbf{k}}=\mathbf{l}_{e}$, as 
measured in  the c.m.\ system, and with scattering angle $\cos \theta=
\hat{\mathbf{k}}\cdot \hat{\mathbf{p}}$. 
 The scattering plane makes up 
the $x'z'$-plane, and with the $y'$-axis  normal to this plane. 
We choose a right-handed-coordinate system with basis vectors
\begin{eqnarray}
	\mathbf{e}'_z  &=&  \hat{\mathbf{k}}, \nonumber \\
	\mathbf{e}'_y  &=& \frac{1}{\sin\theta } ( \hat{\mathbf{k}}\times \hat{\mathbf{p}} ) ,
	\nonumber \\
	\mathbf{e}'_x  &=& \frac{1}{\sin\theta } (\hat{\mathbf{k}}\times \hat{\mathbf{p}} ) 
	 \times\hat{\mathbf{k}}.\label{zaunity}
\end{eqnarray}
In the helicity-coordinate system the final-state-hyperon 
momentum can be  decomposed as
\begin{equation}
	\hat{\mathbf{p}}= \sin\theta\,  \mathbf{e}'_x +\cos\theta\,	\mathbf{e}'_z  ,
\end{equation}
and  ${\mathbf{N}}=-\mathbf{e}'_y $ normal to the scattering plane, 
for ${\mathbf{N}}$
 defined in Eq.(\ref{Ndef}). 

In spherical $x'y'z'$ coordinates the unit vectors $\mathbf{l}_{\Lambda}$ 
and $\mathbf{l}_{\bar{\Lambda}}$
associated with  the directions of motion of 
the $\Lambda$ and $\bar{\Lambda}$ hyperons   are,
\begin{eqnarray}
	\mathbf{l}_\Lambda &=&(\cos \phi'_\Lambda \sin \theta'_\Lambda,  \sin \phi'_\Lambda \sin \theta'_\Lambda, \cos \theta'_\Lambda),\nonumber \\
\mathbf{l}_{\bar{\Lambda}}&=&(\cos \phi'_{\bar{\Lambda}}
	\sin \theta'_{\bar{\Lambda}},  \sin \phi'_{\bar{\Lambda}}
	\sin \theta'_{\bar{\Lambda}}, \cos \theta'_{\bar{\Lambda}}),
	\label{EP-def}
\end{eqnarray}
and similarly for the unit vectors $\mathbf{l}_p$ and
 $\mathbf{l}_{\bar{p}}$.

As in the previous section we introduce a short-hand notation 
for   vectors expressed in helicity coordinates,
 $\mathbf{l}_{\Lambda} =\mathbf{E}'=(E_x', E_y', E_z')$ 
and $\mathbf{l}_{\bar{\Lambda}} =\mathbf{F}'=(F_x', F_y', F_z').$ 
In order to determine the spin-density-distribution function in terms of 
the helicity 
angles we need the six kinematic functions
 $Q(\mathbf{l}_{\Lambda},\mathbf{l}_{\bar{\Lambda}} )$ 
of Eq.(\ref{SixQ}) in terms of the helicity angles of Eqs.(\ref{EP-def}). 
In principle, this is straightforward but it turns out to be more involved than for the global case, since some of the 
$Q(\mathbf{l}_{\Lambda},\mathbf{l}_{\bar{\Lambda}} )$ functions will
depend on the scattering angle $\theta$. 
 
 The basis vectors of Eqs.(\ref{zaunity}) and (\ref{xaunity}) are
 related by
\begin{eqnarray}
\mathbf{e}_x  &=& -\cos \theta\,  \mathbf{e}'_x +\sin \theta\, \mathbf{e}'_z,
	 \nonumber \\
	\mathbf{e}_y  &=& -\mathbf{e}'_y ,
	\nonumber \\
	\mathbf{e}_z  &=& \sin \theta \,\mathbf{e}'_x +\cos \theta\, \mathbf{e}'_z. 
	 \label{zxunity}
\end{eqnarray}
 From this relation  one derives a corresponding relation for the global-vector components $F_k$, and helicity-vector 
components $F_k'$,  of 
the directional unit vector $\mathbf{l}_ {\bar{\Lambda}}$ associated  
with the $\bar{\Lambda}$ hyperon, 
\begin{eqnarray}
	F_x &=& - \cos \theta F_x' + \sin\theta F_z' , \nonumber \\
	F_y  &=& -F_y',
	\nonumber \\
	F_z  &=& \sin\theta F_x'+ \cos\theta F_z' ,\label{Ftran}
\end{eqnarray}
and ditto for the $\Lambda$ hyperon case.

The new set of the six 
 $Q(\mathbf{l}_{\Lambda},\mathbf{l}_{\bar{\Lambda}} )$ functions 
of Eq.(\ref{SixQ}) is obtained by replacing global-vector components 
by helicity-vector components, to give
\begin{align}
	Q_\Lambda =& -E_y',\nonumber\\
	Q_{{\bar{\Lambda}}} =& - F_y'	,\nonumber\\
	Q_1 =&(\sin\theta E_x' + \cos \theta E_z')  
  ( \sin\theta F_x' +\cos \theta F_z') ,
  \nonumber\\
	Q_2 = & Q_3 +  E_y'  F_y', \nonumber \\
Q_3 =& (-\cos\theta E_x ' + \sin \theta E_z') 
  ( -\cos\theta F_x' +\sin \theta F_z'),
		\nonumber\\
	Q_4 =&  (-\cos\theta E_x'+\sin \theta E_z')
	        (\sin \theta F_x' + \cos \theta  F_z' ) \nonumber \\
			&	+ (\sin \theta E_x' +\cos \theta  E_z' ) 
				 (-\cos\theta F_x' +\sin \theta F_z' ). \label{Qprim}            
\end{align}

This set of helicity-angular-dependent functions has a decidedly more complex  dependence  on the scattering angle $\theta$ than the global-angular set of Eq.(\ref{SixEE}), which is independent of the 
scattering angle.  Helicity coordinates are e.g.~used by the BES group,
 \cite{Ablikim17a,Nature}, and  by \cite{CZ}.

The differential-spin-density distribution is defined in 
Eq.(\ref{Weqn:pdf}). For the application to helicity 
coordinates it takes the form 
\begin{equation}
\begin{split}
{\cal{W}}({\boldsymbol{\xi}}(\Omega'))=&\ 1+\eta_\psi\cos^2\! \theta 
     \\ 
		& +\sqrt{1-\eta_\psi^2}\sin(\Delta \Phi_\psi)\sin\theta \cos\theta
		\bigg[\lambda_{\Lambda}E_y' + \lambda_{\bar{\Lambda}} F_y' \bigg]      \\
		&    +\lambda_{\Lambda} \lambda_{\bar{\Lambda}} 
	\bigg[ (1+\eta_\psi ) Q_1
	+\sin^2\! \theta \bigg( (Q_3-Q_1)+\eta_\psi (Q_3-Q_2)\bigg)
	\\ &+
	\  \sqrt{1-{{\eta_\psi}}^2}\cos({{\Delta\Phi_\psi}})\sin\theta \cos\theta\,  
	   Q_4 \bigg], \label{eqn:GKprim}
\end{split}
\end{equation}
with the $Q$ functions as defined in Eqs.(\ref{Qprim}). 
The argument ${\boldsymbol{\xi}}(\Omega')$  of 
the function ${\cal{W}}({\boldsymbol{\xi}}(\Omega'))$ 
remind us 
we work in the helicity-coordinate system.

%
\section{Cross-section distributions}
In view of its simplicity, we propose evaluating the 
cross-section distribution for each event in the global $xyz$ coordinate 
system of Eq.(\ref{xaunity}). 
The expression for the differential-spin-density distribution 
${\cal{W}}({\boldsymbol{\xi}}(\Omega))$ in this
coordinate system is already known, and displayed in Eq.(\ref{eqn:GK}),
 where the symbol $\Omega$ 
refers to spherical angles, $\Omega=(\phi,\theta)$, in the 
$xyz$ coordinate system.


It might be remembered we introduced the notation  
$\mathbf{E}=\mathbf{l}_\Lambda$ and 
$\mathbf{F}=\mathbf{l}_{\bar{\Lambda}}$, with Cartesian components 
as defined in Eq.(\ref{E-def}). 
A unit vector such as $\mathbf{l}_\Lambda$, which is  a unit vector 
in the direction of motion of the $\Lambda$ hyperon in the rest 
system of the $\Sigma^0$ hyperon,   can be expressed in either 
Cartesian $xyz$ or spherical-angular variables,  
\begin{equation}
	\mathbf{l}_{\Lambda}=(l_{\Lambda x}, l_{\Lambda y},l_{\Lambda z})
	 =(\cos \phi_{\Lambda}\sin \theta_{\Lambda}, 
	\sin \phi_{\Lambda} \sin \theta_{\Lambda}, \cos \theta_{\Lambda}).
\end{equation}
The decomposition into spherical coordinates needs to be known 
since the phase-space element $\rd \Omega_\Lambda$ is expressed in 
termes spherical-angular variables.

It was already noticed in Sect.\ref{Sec.Dsb} that the angular variables 
 $\Omega_p$ and 
$\Omega_{\bar{p}}$ only appear in the multiplicative parameters  
$\lambda_\Lambda(\theta_{\Lambda p})$ and 
$\lambda_{\bar{\Lambda}}(\theta_{\bar{\Lambda}\bar{p}}) $ 
of Eqs.(\ref{Ang1}) and (\ref{Ang2}). Averages over 
$\Omega_p$ and $\Omega_{\bar{p}}$ give
\begin{align}
	\left\langle \lambda_\Lambda(\theta_{\Lambda p})\right\rangle
	=&\int \frac{\rd \Omega_p}{4\pi}\ \alpha_\Lambda \cos(\theta_{\Lambda p})
	=\third \ \alpha_\Lambda, \\ 
	\left\langle \lambda_{\bar{\Lambda}}(\theta_{\bar{\Lambda}\bar{p}})\right\rangle=&	\int \frac{\rd \Omega_{\bar{p}}}{4\pi}
	\ \alpha_{\bar{\Lambda} }  \cos(\theta_{\bar{\Lambda}\bar{p}})
	=\third \ \alpha_{\bar{\Lambda} } ,
\end{align}
where $\alpha_\Lambda $ and $\alpha_{\bar{\Lambda} }$ are the
 weak-interaction-decay parameters for the $\Lambda$ 
and $\bar{\Lambda}$ hyperons.

Thus ends  our exposition of the factors making up the
master formula, Eq.(\ref{MasterForm}), for the
normalized cross-section distribution for production and 
decay of $\Sigma^0 \bar{\Sigma}^0$ pairs in $e^+e^-$
annihilation.
\section{Summary}

This is a study of joint production and simultaneous 
sequential decay of $\Sigma^0\bar{\Sigma}^0$ pairs produced in 
$e^+e^-$ annihilation.      
It starts from a  master formula which is a product of 
three factors, describing: the annihilation of a lepton pair 
into a hyperon pair; the spin-density distribution 
${\cal{W}({\boldsymbol{\xi}})}$ representing the spin dependence 
in hyperon production and decay;
  and the phase-space element in  sequential hyperon decay.
 Each measured  event is specified by a nine-dimensional vector 
${\boldsymbol{\xi}}=(\theta,\Omega_{\Lambda},\Omega_{p},\Omega_{\bar{\Lambda}},\Omega_{\bar{p}}),$
with $\theta$ the scattering angle 
in the $e^+e^-\rightarrow \Sigma^0\bar{\Sigma}^0$ subprocess. 

The dynamics of the process is described by four unit-three vectors 
$\mathbf{l}_p,\  \mathbf{l}_{\Lambda} ,\ \mathbf{l}_{\bar{p}},
\ \mathbf{l}_{\bar{\Lambda}}$, directed along the directions
of motion of the final state baryons
$(\Omega_{p},\Omega_{\Lambda},\Omega_{\bar{p}},\Omega_{\bar{\Lambda}})$ .
We have arranged so that the spin-density-distribution function can 
be written as
\begin{equation}
\begin{split}
{\cal{W}}({\boldsymbol{\xi}})=&\ {\cal{R}}- 
		\bigg[ \lambda_\Lambda{{Q}}_\Lambda + 
		\lambda_{\bar{\Lambda}}Q_{\bar{\Lambda}}
		\bigg] {\cal{S}}       \\&             
   +\lambda_\Lambda\lambda_{\bar{\Lambda}}\bigg[  Q_1 {\cal{T}}_1 +
    Q_2 {\cal{T}}_2 + Q_3 {\cal{T}}_3+ Q_4 {\cal{T}}_4 \bigg]
    . \label{WeqnS:pdf}
\end{split}
\end{equation}
Here, the six functions ${\cal{R}}$, ${\cal{S}}$, and ${\cal{T}}$ 
are functions of the scattering angle $\theta$ and the 
ratio of form factors $\eta_\psi$, whereas the 
 six functions $Q$  are functions of  
$ \mathbf{l}_{\Lambda}$ and $ \mathbf{l}_{\bar{\Lambda}}$, 
and of $\hat{\mathbf{p}}=\mathbf{l}_{\Sigma^0}$  and 
$\hat{\mathbf{k}}=\mathbf{l}_{e}$.
The unit vectors
 $\mathbf{l}_{{p}}$ and $\mathbf{l}_{{\bar{p}}}$
only enter the weak-asymmetry functions $\lambda_\Lambda$ and
 $\lambda_{\bar{\Lambda}}$
of Eqs.(\ref{Ang1}) and (\ref{Ang2}).

It remains to connect the four kinematic unit vectors  to measured 
quantities. To this end we imbed   Cartesian-coordinate systems in  
our events. Then, with the Lambda hyperon as an example, 
\begin{equation}
	\mathbf{l}_\Lambda =(l_{\Lambda x}, l_{\Lambda y},l_{\Lambda z} )=(\cos \phi_\Lambda \sin \theta_\Lambda,  \sin \phi_\Lambda \sin \theta_\Lambda, \cos \theta_\Lambda).
\end{equation}
Our preferred coordinate system is named global and has the $xz$-plane 
as scattering plane, and $\hat{\mathbf{p}}$ along the $z$-direction.
In global coordinates the building blocks of the spin-density-distribution function 
${\cal{W}({\boldsymbol{\xi}})}$ in Eq.(\ref{WeqnS:pdf}) 
have the simple structure mentioned above. In particular, the six $Q$ 
functions are independent of the scattering angle $\theta$.

An alternative to global coordinates is helicity coordinates, when
 the $x'z'$-plane is the scattering plane, and $\mathbf{k}$
directed along the $z'$ axis. Several of the $Q$ function now depend 
on the scattering angle $\theta$ in a complex way, even though 
the two coordinate systems are related by a rotation.

%
\appendix
%
%
%
\section{Baryon form factors }\label{tvaa}
%
%

The diagram in Fig.1 describes the annihilation reaction
 $e^-(k_1)e^+(k_2)\rightarrow Y(p_1)\bar{Y}(p_2)$ and involves two 
vertex functions; one of them leptonic, the other one baryonic. The strength of the lepton-vertex function 
is determined by the fine-structure constant $\alpha_e$, but 
two complex form factors  $G^\psi_M(s)$ and $G^\psi_E(s)$ are needed
 for a proper parametrization of   the baryonic vertex function, 
as of Ref.\cite{GF3}. 
The values of these form factors vary with energy,  $s=(p_1+p_2)^2$.

The strength of the baryon form factors is measured
 by the function $D_\psi(s)$, 
\begin{equation}
	D_\psi(s)=s\left| G^\psi_M \right|^2 + 4 M^2 \left| G^\psi_E \right|^2, \label{DS_def}
\end{equation}
with the $M$-variable representing the hyperon mass.  The ratio of  form factors
is measured by $\eta_\psi(s)$,
\begin{equation}
	\eta_\psi(s)= \frac{s\left| G^\psi_M \right|^2 - 4 M^2\left| G^\psi_E \right|^2}
	{s\left| G^\psi_M\right|^2 + 4 M^2\left| G^\psi_E\right|^2},\label{alfa_def}
\end{equation}
with $\eta_\psi(s)$ satisfying $-1\leq \eta_\psi(s) \leq 1$.
The relative phase of form factors is measured by $\Delta\Phi_\psi(s)$, 
\begin{equation}
	\frac{G^\psi_E}{G^\psi_M}=e^{i\Delta\Phi_\psi(s)} \left| \frac{G^\psi_E}{G^\psi_M}\right|. \label{DPHI_def}
\end{equation}

%
%
\section{Structure functions}\label{AppC}
%

The six structure functions ${\cal{R}}$, ${\cal{S}}$, and ${\cal{T}}$ of
 Eq.(\ref{ISD})  
 depend on the scattering angle $\theta$, the ratio function $\eta_\psi(s)$, and
the phase function $\Delta\Phi_\psi(s)$. To be specific \cite{GF2,GF3},
\begin{eqnarray}
{\cal{R}} &=& 1 +\eta_\psi  \cos^2\!\theta, \label{DefR}\\
  {\cal{S}} &=& \sqrt{1-\eta_\psi^2}\sin\theta\cos\theta\sin(\Delta\Phi_\psi), \label{DefS}\\
	{\cal{T}}_1 &=& \eta_\psi + \cos^2\!\theta, \\
	{\cal{T}}_2 &=& -\eta_\psi\sin^2\!\theta,  \\
	{\cal{T}}_3 &=&( 1+\eta_\psi) \sin^2\!\theta , \label{NewT3}\\
	{\cal{T}}_4 &=& \sqrt{1-\eta_\psi^2}\sin\theta\cos\theta\cos(\Delta\Phi_\psi). \label{RSTdef}
\end{eqnarray}
The parameters $\eta_\psi$ and $\Delta\Phi_\psi$ are defined in Eqs.(\ref{alfa_def}) and (\ref{DPHI_def}). The function ${\cal{T}}_3$ 
of Eq.(\ref{NewT3}) differs from the corresponding function 
${\cal{T}}_3$ of Ref.\cite{GKa} by the $\sin^2\!\theta $ factor. Similarly, 
the function ${\cal{T}}_4$ 
of Eq.(\ref{RSTdef}) differs from the corresponding 
function ${\cal{T}}_4$ of 
Ref.\cite{GKa} by the $\sin\theta $ factor.

%
\section{Finding angular variables}\label{sex}


The angular functions 
$Q(\mathbf{l}_{\Lambda},\mathbf{l}_{\bar{\Lambda}})$  
of Eq.(\ref{SixQ}) and the $\lambda$ parameters of Eqs.(\ref{Ang1}) 
and (\ref{Ang2}) 
are expressed in terms of unit vectors such as $\mathbf{l}_p$
and $\mathbf{l}_{\Lambda}$, which are not directly measurable but which
must be calculated. We suggest the following approach.

For each event we imbed  the particle momenta in its c.m.\ system and with 
coordinate axes as defined in Eq.(\ref{xaunity}).   For the $\Sigma^0$ hyperon the
components of the momentum are, by definition,
\begin{equation}
	\hat{\mathbf{p}}_{\Sigma^0}=(0, 0,1).
\end{equation}

Then, let us consider the  proton and the hyperon of the final state, 
with momenta $\mathbf{p}_p$ and $\mathbf{p}_\Lambda$  
in the c.m.\ system. In the rest system 
of the Lambda hyperon, $\mathbf{L}_p$ denotes the proton momentum, 
which is given by the expression 
\begin{align}
	 \mathbf{L}_p &= 
	\mathbf{p}_p +B_{\Lambda p} \mathbf{p}_{\Lambda} , \label{pcm} \\
	B_{\Lambda p}&=\frac{1}{m_\Lambda}
	\bigg[ \frac{1}{E_\Lambda+m_\Lambda}\, \mathbf{p}_{\Lambda}\cdot \mathbf{p}_{p} 
	  - E_\Lambda  \bigg]. 
\end{align}
Now, the length of the vector $\mathbf{L}_p$ is well-known,  being the  momentum 
in the hyperon decay $\Lambda\rightarrow \pi N $, and therefore
\begin{equation}
	| \mathbf{L}_p |=\frac{1}{2m_\Lambda}
	\bigg[ ( m_\Lambda^2 +    m_\pi^2 -  m_N^2 )^2  -4 m_\Lambda^2 m_\pi^2 \bigg]^{1/2}.
	 \label{Lpvec}
\end{equation}
Hence, the unit vector $\mathbf{l}_p$ appearing in our  equations should be
\begin{eqnarray}
	\mathbf{l}_p&=& \mathbf{L}_p/| \mathbf{L}_p |, \\
	 &=&(\cos \phi_ p\sin \theta_p,  \sin \phi_p \sin \theta_p, \cos \theta_p).
\end{eqnarray}

  Also, the equation for $\mathbf{l}_\Lambda$ in the decay $\Sigma^0\rightarrow\Lambda\gamma$ is easily written down, as are the corresponding 
	equations for the antiparticles,  $\bar{{p}}$ and  $\bar{{\Lambda}}$.

\section*{Acknowledgments}
I would like to thank Karin Sch{\"o}nning for informative discussions
%


\begin{thebibliography}{99}

\bibitem{Ablikim17a}  M.~Ablikim { et~al.}~(BESIII), Phys.\ Rev.\ D  {\bf 95},  052003 (2017).
\bibitem{GKa} G.\ F\"aldt and K. Sch{\"o}nning, Phys.\ Rev.\ D   {\bf 101},  033001 (2020).
\bibitem{Nature} M.~Ablikim et al.\ (BESIII), 
   Nat.~Phys.~{\bf 15},~631 (2019) 
\bibitem{GF3} G.~F\"aldt and A.~Kupsc, Phys.\ Lett.\ B {\bf 772},  16 (2017).
\bibitem{GF1} G.~F\"aldt, Eur.\ Phys.\ J.\ A  {\bf 51},  74 (2015).
\bibitem{GF2} G.~F\"aldt, Eur.\ Phys.\ J.\ A  {\bf 52},  141 (2016).
\bibitem{GFjuni} G.~F\"aldt, \textit{Sequential hyperon decays} (Lecture notes,
    Uppsala, June 2017).
\bibitem{Lee} T.\ D.\ Lee and C.\ N.\ Yang,  Phys.\ Rev.\    {\bf 108},  1645 (1957).
\bibitem{Don1} John F.\ Donoghue and Sandip Pakvasa, Phys.\ Rev.\ Lett.\ {\bf 55},  162 (1985).
\bibitem{Don2} John F.\ Donoghue, Xiao-Gang He, and Sandip Pakvasa, Phys.\ Rev.\ D{\bf 34},  833 (1986).
\bibitem{CZ} H.\ Czy\.{z}, A.\ Grzeli\'{n}ska, and J.H.\ K\"{u}hn, Phys.\  
  Rev.\ D{\bf 75}, 074026 (2007).
\end{thebibliography}
\end{document}